\documentclass[12pt,preprint]{aastex}

\shorttitle{Resolution of Distance Ambiguities}
\shortauthors{Sewilo et al.}

\received{2004 January 27}
\begin{document}

\title{Resolution of Distance Ambiguities of Inner Galaxy Massive Star Formation Regions II}

\author{M. Sewilo\altaffilmark{1}, C. Watson\altaffilmark{1}, E. Araya\altaffilmark{2}, E. Churchwell\altaffilmark{1}, P. Hofner\altaffilmark{2,3}, and  S. Kurtz \altaffilmark{4}} 

\altaffiltext{1}{University of Wisconsin - Madison, Department of Astronomy, 475
N. Charter St., Madison, WI 53706}

\altaffiltext{2}{New Mexico Tech, Physics Department, 801 Leroy Place, Socorro, NM 87801}

\altaffiltext{3}{National Radio Astronomy Observatory, PO Box 0, Socorro, NM 87801}

\altaffiltext{4}{Centro de Radioastronom\'\i a y Astrof\'\i sica, UNAM, Apdo. Postal 3-72, 58089, Morelia, Mich. Mexico}

\begin{abstract}
We report simultaneous H110$\alpha$ and H$_{2}$CO line observations with the NRAO
Green Bank Telescope toward 72 H {\small II} regions in the SPITZER/GLIMPSE survey
area ($|$l$|$ = 10$\arcdeg$ $-$ 65$\arcdeg$ and $|$b$|$ $\le$ 1$\arcdeg$). We used the H110$\alpha$ line to
establish the velocity of the H {\small II} regions and H$_{2}$CO absorption lines to distinguish between near and far distances. Accurate
distances are crucial for the determination of physical properties of massive star formation
regions. We resolved the distance ambiguity of 44 H {\small II} regions.  We
detected multiple H {\small II} regions along 18 lines of sight located in the
longitude interval 12$\arcdeg$ to 31$\arcdeg$, primarily a result of the relatively large telescope
beam width.  We could not resolve distance ambiguities for lines of sight
with multiple H {\small II} regions, since we could not determine which H$_{2}$CO lines
were being absorbed against which H {\small II} region.

We examined the projected location of H {\small II} regions whose distance
ambiguities have been resolved (in this work and other similar studies)
in the Galactic plane and in a longitude-velocity diagram for a
recognizable spiral arm pattern. Although the highest density of points
in the position-position plot approximately follows the spiral arms
proposed by Taylor and Cordes (1993), the dispersion is still about as
large as the separation between their proposed arms. The
longitude-velocity plot shows an increase in the density of sources at the points where the
spiral arm loci proposed by Taylor and Cordes (1993) are approaching  the locus of tangent point velocities
and a lower density between the arm loci. However, it is not possible to trace spiral arms over
significant segments of Galactic longitude in the longitude-velocity
plot. We conclude that a very large number of H {\small II} regions in combination
with more sophisticated Galactic rotation models will be required to
obtain a more continuous spiral pattern from kinematic studies of
H {\small II} regions than from fully sampled surveys of H {\small I} or CO.
\end{abstract}


\section{INTRODUCTION}
The determination of distances to astronomical objects is perhaps one 
of the oldest and most important problems in astronomy. Accurate distances are necessary to establish physical properties of sources as well as the distribution of different classes of objects in the Galaxy. 
In particular, for objects that are believed to be tracers of spiral structure, such as massive star formation regions, 
one may infer the number and location of spiral arms in the Galaxy. For most sources
located in the Galactic plane, optical distance determination methods 
cannot be used due to the large opacity of interstellar dust. The usual way to estimate distances to radio sources (e.g., molecular clouds, H {\small II} regions, PNs) is based on the kinematic properties of the Galactic plane. The velocity field of the Galaxy can be approximated by axially-symmetric circular orbits in which the velocity is only a function of Galactocentric distance (e.g., Brand \& Blitz 1993, and references therein). The distance to an object in the Galactic plane can be obtained from its observed radial velocity with respect to the local standard of rest (LSR) reference frame by inversion of the mathematical relationship between the velocity and Galactocentric distance (i.e., rotation curve). This procedure gives a unique solution for the kinematic distance (for a given longitude and radial velocity) in the outer Galaxy, however the LSR distance is degenerate in the inner Galaxy (quadrants I and IV). In the inner Galaxy, the line of sight to a given object intersects its circular orbit at two points that define the {\it near} and {\it far} positions. At these two locations an object has the same radial velocity, giving rise to a kinematic distance degeneracy which is referred to as {\it the distance ambiguity problem}. Only inner Galaxy sources located at the {\it tangent point} have unambiguous distances. The tangent point is the point at which a line of sight to an object is tangent to its orbit. 

This work is the third in a series (Araya et al. 2002, Paper I; Watson et al. 2003, Paper II) intended to resolve the distance ambiguity of bright H {\small II} regions located within the area of 
the SPITZER/GLIMPSE survey of the inner Galaxy ($|$l$|$ = 10$\arcdeg$ $-$ 65$\arcdeg$, $|$b$| \le$ 1$\arcdeg$). The motivation for this work 
stems from the fact that: (a) accurate measurements of distances to H {\small II} regions and their associated star forming regions are necessary to determine their bolometric luminosity, mass, and size. The physical parameters of massive star forming regions constrain theoretical models of star formation and evolution of molecular clouds; and, (b) regions of massive star formation are expected to trace the spiral arms of the Galaxy and
therefore, the establishment of distances to these objects provides an opportunity to improve our understanding of the large-scale structure of the Galaxy.

In the first part of this project, 75 H {\small II} regions with Galactic longitudes ranging from $+$30$\arcdeg$ to $+$70$\arcdeg$ and Galactic latitude between $-$1$\arcdeg$ and $+$1$\arcdeg$, were observed in two surveys with the Arecibo 305 m telescope (the half-power beamwidth $\sim$ 1$'$; Paper I and Paper II). The Arecibo antenna was used to simultaneously observe the H110$\alpha$ line in emission and the $F_{Ka Kc} = 1_{10}-1_{11}$ line of H${_2}$CO in absorption against H {\small II} regions. The H110$\alpha$ line was used to establish the velocity of the H {\small II} regions and the H${_2}$CO absorption components were used to distinguish between near and far distances. It was assumed that all H$_{2}$CO absorption lines are produced by absorption of radio continuum from a single background H {\small II} region, thus neglecting  absorption of the cosmic microwave background radiation. This assumption is supported by the fact that only seven H$_{2}$CO absorption features were observed in off-source integrations, whereas 134 were detected in on-source spectra. The rotation curve of Brand and Blitz (1993) was used to derive kinematic distances. Distance ambiguities were resolved for $\sim$95$\%$ (20 sources) and $\sim$81$\%$ (44 sources) of the sample in the first (Paper I) and second (Paper II) Arecibo surveys, respectively. Here we report observations of selected  H {\small II} regions in the northern SPITZER/GLIMPSE survey area that is not accessible from Arecibo. The observations were made using the National Radio Astronomy Observatory\footnote{The National Radio Astronomy Observatory is a facility of the National Science Foundation operated under cooperative agreement by Associated Universities, Inc.} (NRAO) 100 m Green Bank Telescope (GBT).  Distance ambiguities are resolved using the same method as in the Arecibo surveys. A longitude-velocity diagram based on the Arecibo and GBT surveys is constructed.

\section{OBSERVATIONS}

The observations were made on 2002 November 1$-$3, 8$-$9, and 14 with the GBT. Seventy two sources were selected from the IRAS Point Source Catalog that satisfy the far-infrared color-color criteria of UC H {\small II} regions (i.e., log(S$_{60\mu m}$/S$_{12\mu m}$)$\ge$ 1.30 and log(S$_{25\mu m}$/S$_{12\mu m}$) $\ge$ 0.57; Wood and Churchwell 1989), and have 100 $\mu$m flux densities $\ge$ 1000 Jy. In addition, we required that all candidate sources, visible with the GBT, lie within Galactic longitude ranges, $-$20$\arcdeg$ to $-$10$\arcdeg$ and $+$10$\arcdeg$ to $+$30$\arcdeg$, and have Galactic latitudes between $-$1$\arcdeg$ and $+$1$\arcdeg$. These areas are included in the SPITZER/GLIMPSE survey region ($|$l$|$= 10$^\circ$$-$65$^\circ$ and $|$b$|$ $\le$ 1$^\circ$) and were not sampled by the Arecibo surveys. The observed sources are listed in Table 1. The main goal of this survey is to establish reliable distances to massive star formation regions (young OB associations and clusters) that will be observed at mid-infrared wavelengths by the GLIMPSE Legacy Science Program with the Spitzer Space Telescope. The GBT observations supplement the Arecibo data which together cover the SPITZER/GLIMPSE survey area accesible from the northern hemisphere.

The 1$_{10}$-1$_{11}$ transition of formaldehyde (H${_2}$CO, $\nu_{o}$ = 4829.6594 MHz) was observed in absorption against H {\small II} 
regions. Simultaneously, the H110$\alpha$ ($\nu_{o}$ = 4874.1570 MHz) radio recombination line was observed to establish kinematic 
distances. Both spectral lines were measured using the C-band Gregorian focus receiver. The half-power beamwidth (HPBW)
of the GBT at 4.86 GHz was measured to be 2$'$.56 $\pm$ 0$'$.12. The 4 IF mode of the correlator was used to observe two transitions simultaneously in 
both senses of linear polarization with two 
subcorrelators centered on the H${_2}$CO line, and the other two centered on the H110$\alpha$ radio 
recombination line. Each subcorrelator had 4096 channels, resulting in a resolution of 3.05 kHz (0.188 km s$^{-1}$), a total bandwidth of 12.5 MHz, 
and a total velocity coverage of $\sim$770 km s$^{-1}$. The central velocity for each subcorrelator was set to V$_{LSR}$ = 0 
km s$^{-1}$. The correlator used 9-level sampling.

Position switching was used with a noise-diode calibration signal injected into the system every second. 
The OFF position was chosen to trace the same path on the sky as the ON$-$source position. Depending on the continuum flux densities, the observations required 5 to 30 minutes ON$-$source integration time (10$-$60 minutes/source) to achieve 
a signal to noise $\ge$ 5 (see Table 1). For both polarizations, the system temperature during the 
observations was 26 $\pm$ 2 K. Observations of the flux calibrator NGC 7027 with assumed flux density of 5.37 Jy at 4.82 GHz 
indicate that the antenna gain remained constant at 1.83 K Jy$^{-1}$, independent of zenith and azimuth angles (11$\arcdeg$.6 $-$ 56$\arcdeg$.5 and $-$68$\arcdeg$.8 $-$ $+$70$\arcdeg$.3, respectively). 
For each source both ON and OFF positions were inspected individually to check for polarization, absorption and/or emission 
in the OFF position, and radio interference. The final spectra were obtained by averaging both 
polarizations and all scans for each source.

\section{RESULTS}

The data were calibrated using the Astronomical Imaging Processing System (AIPS++)\footnote{AIPS++ is a software package of the National Radio Astronomy Observatory (NRAO)}. 
Subsequent spectral line analysis used the CLASS software package\footnote{CLASS is part of the GLIDAS software package developed by the Institut de Radio Astronomie Millim\'etrique (IRAM).}. The H110$\alpha$ and H$_2$CO line parameters were derived by fitting Gaussian profiles to each velocity component along the line of sight. Before the line parameters were measured, all spectra were Hanning smoothed to the velocity resolution listed in Table 2.  Continuum emission was detected in all but one source (G345.40$-$0.94). The results of the spectral line analysis are presented in Tables 3, 4, and 5 for the sources with single H110$\alpha$ velocity components along a given line of sight, multiple H110$\alpha$ velocity components (see discussion below), and H${_2}$CO absorption lines, respectively. The full-widths at half-power (FWHP) of the spectral lines are corrected for broadening introduced by the spectrometer and smoothing. The errors quoted in Tables 3$-$5 are the formal 1$\sigma$ errors (68.3 \% confidence level) of the Gaussian fits to the data. Only the spectral lines with intensities greater than 3$\sigma$ were included in the tables. Figures 1.1$-$1.72 show both the H110$\alpha$ emission and H${_2}$CO absorption spectra for each source in our sample. Flux densities throughout the paper are given in Jy.

Seventy two massive star formation regions were observed. We detected formaldehyde absorption toward all the sources, and H110$\alpha$ emission toward all but 6 sources (G340.05$-$0.24, G345.00$-$0.22, G345.40$-$0.94, G345.50$+$0.35, G12.02$-$0.03, and G12.89$+$0.49). Eighteen sight lines had multiple H110$\alpha$ velocity components (see Table 4), all of which are located in the longitude interval 12$\arcdeg$ to 31$\arcdeg$. Extensive tests of the local oscillator and the system were made, including observations of several lines of sight, to confirm that indeed the observed spectra are repeatable, thus faithful representations of multiple H {\small II} regions along these lines of sight and not instrumental artifacts.  Injection of a test tone into the system did not show the multiple line effect. We also shifted the local-oscillator frequencies LO1 and LO2 by 1 MHz, and did not measure any change in the spacing of the two (or three) lines.  We did not detect multiple H110$\alpha$ radio recombination lines in our Arecibo observations. This non-detection is probably primarily the result of the beam area of the GBT ($\Omega_{A} \sim 2.5 \times 10^{-3}$ deg$^{2}$; HPBW$\sim$2'.6) being seven times larger than the beam area of the Arecibo telescope ($\Omega_{A} \sim 3.7 \times 10^{-4}$ deg$^{2}$; HPBW$\sim$1'), so the probability of having more than one H {\small II} region in a beam is much higher.

\subsection{Distance Classification}

For the inner Galaxy (quadrants I and IV) H {\small II} regions  with a H110$\alpha$ line detection, the near/far distance classification was based on the presence of H$_{2}$CO absorption lines between the source velocity ($|V_{HII}|$) plus 10~km~s$^{-1}$ and the tangential velocity (velocity at the tangent point; hereafter $|V_{TP}|$) plus 10~km~s$^{-1}$. If the H$_{2}$CO absorption feature was found in this velocity range, we classified the source at the far distance; otherwise, the near distance was assigned. The addition of 10~km~s$^{-1}$ to the source and tangent point velocities account for typical departures from circular motions (Burton 1974). We do not resolve the near/far distance ambiguity for H {\small II} regions with velocities higher than $|V_{TP}|$ $-$ 10~km~s$^{-1}$ and lower than $|V_{TP}|$, because we cannot distinguish between near and far distances in this velocity range.  If the velocity of an H {\small II} region is higher than $|V_{TP}|$, we locate this source at the tangent point.  The excursions of the data points to velocities higher than $|V_{TP}|$ values are similar to the H {\small I} excursions associated with the frequently-discussed ``bumps'' in the rotation curve, first pointed out by Kwee et al. 1954.

The distance ambiguity can also be resolved for the intervening molecular clouds. If the velocity of a H$_{2}$CO absorption feature ($|V_{H_{2}CO}|$) is less than $|V_{HII}|$ $-$ 10~km~s$^{-1}$, we locate a molecular cloud at the near distance. However, if $|V_{H_{2}CO}|$ lies between $|V_{HII}|$ $+$ 10~km~s$^{-1}$ and  $|V_{TP}|$ $-$ 10~km~s$^{-1}$, the distance ambiguity cannot be resolved, because in either case (near or far position) the molecular gas is in front of the H {\small II} region. The velocities of many absorption features lie within 10~km~s$^{-1}$ of $|V_{HII}|$, demonstrating that the H {\small II} regions are closely associated with molecular gas. In this case, we give the same distance classification for both H {\small II} regions and H$_{2}$CO clouds associated with them. As in the case of H {\small II} regions, we do not assign distances to molecular clouds with velocities between $|V_{TP}|$ $-$ 10~km~s$^{-1}$ and $|V_{TP}|$, and we locate them at the tangent point if their velocities are higher than $|V_{TP}|$. A brief summary of the criteria used to assign kinematic distances to H {\small II} regions and H$_{2}$CO clouds is presented in Table 6.   

In Table 7 we report the kinematic parameters of 48 H {\small II} regions
and 149 intervening H$_{2}$CO clouds. H {\small II} regions with no
H110$\alpha$ line detection or multiple H110$\alpha$ lines and H$_{2}$CO
clouds observed in their direction, as well as H$_{2}$CO molecular clouds
with ambiguous distances are not included. 
All distances in Table 7 were calculated assuming the Brand and Blitz (1993) Galactic
rotation model. In Table 7, we report the radial velocities of the sources (V$_{LSR}$), the near (D$_{near}$) and far (D$_{far}$) distances, tangent point velocities (V$_{TP}$), 
distances to the tangent point (D$_{TP}$), the adopted kinematic distances to the sources (D$_{LSR}$), distances from the Galactic plane (z), and 
distances to the Galactic center (R$_{GC}$). The errors of the kinematic distances were calculated assuming a velocity
uncertainty due to peculiar motions of 10~km~s$^{-1}$. We resolved the
distance ambiguity toward 44 out of 48 H {\small II} regions with a single
H110$\alpha$ line velocity.  Fourteen H {\small II} regions are classified
at the near distance, 26 at the far distance, and 4 are placed at the tangent point.
We do not give a distance to 4 H {\small II} regions (see Table 6 for classification criteria).  
We classify 72 out of 149 H$_{2}$CO clouds
listed in Table 7 at the near distance. We do not assign a distance to 7
H$_{2}$CO clouds, and place 5 at the tangent point. 64 molecular clouds
are associated with H {\small II} regions, 16 of which are at the
near distance, 37 at the far distance, 5 at the tangent point, and 6 are not classified. One molecular cloud (MC 2) observed toward
G345.49$+$0.31 has an unambiguous distance, because its velocity is
positive and it lies outside the solar circle.

\subsection{Comparison with Other Studies}

To check our results, we observed two sources from the first (G37.87$-$0.40 and G61.48$+$0.09; Paper I) and two from the second Arecibo surveys (G37.76$-$0.20, G49.21$-$0.35; Paper II). We agree on the near-far distance ambiguity resolution for all four sources. We also observed a total of 11 H {\small II} regions observed by Downes et al. (1980), Kolpak et al. (2003), and Fish et al. (2003). We compared our results with their near/far distance classifications for the common sources. 

Kolpak et al. (2003) and Fish et al. (2003) resolved the distance ambiguity for 49 and 20 H {\small II} regions with known radial velocities, respectively, by measuring the 21 cm H {\small I} absorption spectrum toward each source with the VLA. We share one source (G29.96$-$0.02) with both Kolpak et al. (2003) and Fish et al. (2003). Kolpak et al. (2003) placed this source at the far distance, while Fish et al. (2003) put it at the tangent point. We agree with Kolpak at el. (2003) on the far distance. We share two more sources (G21.87$+$0.01 and G23.43$-$0.21) with Kolpak et al. (2003) and agree on all distance determinations, and one source (G345.00$-$0.22) with Fish et al. (2003) that we do not classify, because we did not detect the radio recombination line toward it. 

Six of the sources from our sample were previously discussed by Downes et al. (1980) who observed H110$\alpha$ emission and H$_{2}$CO absorption lines toward H {\small II} regions with the Effelsberg 100 m telescope. Downes et al. (1980) resolved the distance ambiguity for four out of six sources we have in common. Of these, we agree on one (G19.07$-$0.28) and disagree on two (G10.16$-$0.35 and G10.31$-$0.15). We do not classify the fourth one (G24.68$-$0.16), because multiple recombination lines were detected toward this source. 

The two sources on which we disagree with Downes et al. (1980), G10.16$-$0.35 and G10.31$-$0.15, are part of the W31 complex of bright H {\small II} regions and molecular clouds in the inner Galaxy. Wilson (1974) and Kalberla et al. (1982) detected H$_{2}$CO, OH, and H {\small I} in absorption toward the H {\small II} regions in the W31 complex and found absorption line velocities up to 33~km~s$^{-1}$ greater than that of the H {\small II} regions. They argued, however, that W31 is unlikely to be at the far distance because none of the absorbing clouds (including H {\small I}) achieved velocities as large as the tangent point velocity. They argued instead that W31 with a peculiar velocity of $\sim$30~km~s$^{-1}$ is likely to be located on the near side of the expanding ``3-kpc arm'' at a distance of $\sim$5.5 kpc from the Sun. Corbel et al. (1997) argue against the association of W31 with the 3-kpc arm because their CO observations show that the CO clouds associated with W31 departs from the well defined velocity of the 3-kpc arm by $\sim$23~km~s$^{-1}$ ($\sim$5 times the velocity dispersion of the arm); they also claim that the 3-kpc arm is quite deficient in star formation while W31 contains some of the brightest H {\small II} regions in the inner Galaxy, and the CO cloud MC 13 (associated with W31) is far more massive than any cloud in the 3-kpc arm. Corbel et al. (1997) therefore claim that W31 lies at the far distance. Based on our H$_{2}$CO data and classification criteria, we locate W31 at the far distance; however, non-detection of H {\small I} at the tangent point velocity is problematic.

\section{Discussion}

\subsection{Distribution of H {\small II} Regions in the Galactic Plane}

When viewed at optical wavelengths, the arms in spiral galaxies are traced by young massive stars 
and their H {\small II} regions. A compelling explanation for this is that compression due to spiral density 
waves initiates star formation near the leading edge of spiral arms (Roberts 1969). Whatever the reason, 
we know empirically that young massive stars are good tracers of spiral structure in galaxies. It is, 
therefore, natural to ask if the kinematically determined distances of H {\small II} regions, whose distance
ambiguities have been resolved, indicate a recognizable spiral arm pattern in the Milky Way.

In Figures 2 and 3 we show two plots to assist in answering this question. Figure 2  shows
the projected location in the Galactic plane of the H {\small II} regions whose distance ambiguities 
have been resolved. In this figure, we view the Galactic plane from the north Galactic pole with
the Galactic center at (0, 0) kpc and the Sun at (0, 8.5) kpc; both are
designated by a filled circle. In all surveys included in this figure,
the rotation constants, the Galactocentric distance of the Sun
(R$_{o}$) and its rotational velocity around the Galactic center
($\Theta_{o}$) were assumed to be 8.5 kpc and 220 km s$^{-1}$ (Kerr
and Lynden-Bell 1986), respectively. In Araya et al. (2002), Watson et
al. (2003) and in the present study distances were derived assuming that
the Galaxy rotates according to the model of Brand and Blitz (1993). 
For sources from Kolpak et al. (2003) and Fish et al. (2003), who used the Clemens (1985) and flat
rotation curves, respectively, all distances were recalculated using the Brand and Blitz (1993) rotation curve for
consistency with the rest of the data. The solid curves in Figure 2 indicate 
the centroids of the four spiral arms proposed initially by Georgelin and Georgelin (1976) 
and modified by Taylor and Cordes (1993; hereafter TC93). The TC93 model is based on the
standard rotation constants (R$_{o}$, $\Theta_{o}$) = (8.5 kpc, 220 km s$^{-1}$). Figure 3 shows a longitude-velocity 
(l$-$v) plot with the same set of H {\small II} regions as in Figure 2. 
We emphasize that the H {\small II} region data points do not depend on any model assumptions about Galactic rotation. 
We have superimposed the locus of points of the four spiral arms and the tangent point velocity from the TC93 model
using the Brand and Blitz (1993) rotation curve for comparison with the distribution of H {\small II} regions.

What can we conclude from these plots? First, no sharp, ``well defined'' spiral pattern is apparent when 
looking at the distribution of H {\small II} regions projected onto the Galactic plane. However, we have to 
remember that spiral arms are not like the thin lines drawn in Figure 2; they have 
significant widths which will cause some smearing even if we could determine positions precisely. 
If there were no curves to guide the eye in Figure 2, one might come close to
designating similar spiral arms as TC93 if the loci of highest density of points are followed. The
preponderance of H {\small II} regions tend to lie closer to the TC93 spiral arm centroids than in the regions 
between the centroids. Exceptions to this are in the following longitude intervals: 10$\arcdeg$$-$
35$\arcdeg$ where a significant number of objects lie between the Perseus and Sagittarius centroids; 
5$\arcdeg$$-$15$\arcdeg$ where a number of sources lie between the Norma and Scutum centroids;
near 0$\arcdeg$ six or so sources clearly lie between the Scutum and Sagittarius centroids. At
least 9 sources appear to be located in the ``Local Spur'' including the Sun from 75$\arcdeg$ $\le$ l 
$\le$ 90$\arcdeg$.

In the first quadrant of the Galactic plane the TC93 Perseus, Sagittarius, and Scutum spiral arm 
centroids have separations $\le$ 2 kpc. In those longitude intervals where this is the case, positions 
would have to be determined to very high accuracy to clearly distinguish between the three arms. Most 
of the sources that lie well between the TC93 arm centroids have longitudes from near 0$\arcdeg$ to 
$\le$ 20$\arcdeg$ where kinematic distances have lower reliability than at higher longitudes because 
radial velocities are dominated by peculiar motions rather than by circular velocities. Kinematic 
distances have an intrinsic uncertainty because they depend on a rotation model of the Galaxy, that
is known to have relatively large departures from the true motions of the Galaxy at some longitudes.
To first order, a circular rotation model fits the observed velocities relatively well, but departures 
from this general pattern, such as radial motions, occur at all longitudes. A good example 
of this is illustrated in Figure 3 where several H {\small II} regions have velocities 
10 to 25 km~s$^{-1}$ in excess of the tangential velocity. It is therefore not surprising that there is 
a lot of scatter in the position-position plot of Figure 2, especially at low longitudes. Until more 
precise Galactic rotation models are available, this may be the limit to which spiral arms can be 
determined using kinematic distances.

The l$-$v plot in Figure 3 is an effort to examine possible spiral structure using 
only observed quantities of H {\small II} regions; the positions of the H {\small II} region and H {\small I} data 
are independent of a Galactic rotation model. We plot the TC93 
spiral arms using the Brand and Blitz (1993) rotation curve and tangential velocity centroids to show where they lie in an l$-$v diagram, the location of 
the observed sources, however, do not depend on a Galactic rotation model to be placed in this diagram. 
Ignoring the TC93 arm and tangential velocity centroids, we see that sources lie in restricted areas
of the l$-$v plane mostly bound by the TC93 tangential velocity centroids. Also the sources are not
uniformly distributed between the TC93 tangential velocity centroids, but are clumped in the northern 
plane around l $\approx$ 30$\arcdeg$ and v $\approx$ 100 km~s$^{-1}$ and
l $\approx$ 40$\arcdeg$$-$45$\arcdeg$ and v $\approx$ 50 km~s$^{-1}$.  In the TC93 model one sees that in the l$-$v diagram these are locations where the spiral arm loci of TC93 are approaching the tangent point velocity locus; one should expect an increase in the number of sources at these locations due to velocity crowding. There is also an obvious significant decrease in the number of sources between those two locations in the northern plane. The density of sources is generally higher in the regions where the TC93 spiral arm loci are
located than away from them, but there are no obvious sharp continuous features that can be
unambiguously traced in the l$-$v diagram that would clearly indicate a spiral arm. Also, we see several
sources that exceed the TC93 tangential velocities by $\ge$ 10 km~s$^{-1}$; however, their velocities are not higher than the observed H {\small I} cut-off velocities 
(i.e., the most extreme velocities on the H {\small I} spectra; see Figure 3). 

Why is a spiral pattern not obvious in the l$-$v plane? This probably has several reasons. One, spiral 
arms have intrinsic widths both in space and velocity which will smear the distribution of sources 
about arm centroids. Second, peculiar motions (i.e., departures from the motions included in the 
rotation model from which distances were determined) will smear out the distribution around a spiral 
arm centroid even further. Third, unlike the measurement of H {\small I} and CO where continuous kinematic
features can be traced over large ranges of Galactic longitude, measurement of a limited number of
discrete H {\small II} regions does not provide the velocity continuity necessary to recognize large scale 
kinematic features. If the spiral arms in the inner Galaxy are only separated by 
$\le$ 2 kpc as implied by the TC93 model, it is unlikely that kinematic distances, with their 
uncertainties due to distance ambiguities and peculiar motions, to a hundred or so discrete sources 
will be capable of delineating spiral arm centroids precisely enough to clearly distinguish between
spiral arms, which is what Figure 2 illustrates.

\section{Summary and Conclusions}

We report simultaneous H110$\alpha$ and H$_{2}$CO line
observations toward 72 H {\small II} regions using the NRAO GBT. The main goal of
this program was to obtain kinematic distances and to resolve the
near/far distance ambiguity for inner Galaxy H {\small II} regions. The main
motivation to do this was to obtain accurate distances to massive star
formation regions in the inner Galaxy in preparation for the
SPITZER/GLIMPSE survey. The H$_{2}$CO 6 cm line was detected in absorption
toward all 72 H {\small II} regions and the H110$\alpha$ line was detected toward all but
6 H {\small II} regions. Most lines of sight had multiple velocity components in
H$_{2}$CO (on average $>$ 3). 

Of the 72 H {\small II} regions observed in the H110$\alpha$ line, 18 in the longitude range 12$\arcdeg$ $\le$ l
$\le$ 31$\arcdeg$ had multiple velocity components; because we could not determine
which lines were being absorbed against which H {\small II} region, we could not
resolve distance ambiguities for these sources. Of the remaining 54 H {\small II}
regions, 6 were not detected in the H110$\alpha$ line. We were able to resolve
the distance ambiguity of 44 out of the remaining 48 H {\small II} regions, 14 of
which are at the near distance, 26 at the far distance, and 4 at the
tangent point. Of 149 H$_{2}$CO clouds (i.e., velocity components), we found that 72 are at the near distance, 
5 are at the tangent point, 7 are not classified, 64 are associated with H {\small II} regions, and one is located
outside the solar circle (G345.49$+$0.31 MC 2).
Sixteen out of 64 molecular clouds associated with H {\small II} regions are at the near distance, 37 are at
the far distance, 5 are at the tangent point, and 6 are not classified. 

In those cases where we have sources in common
with other groups who have resolved distance ambiguities, we generally
agree on distance assignments with only a few exceptions that mostly
arise from differences in the range of peculiar motions considered. 

The sources with resolved distance ambiguities
were plotted in position-position and l$-$v
plots to determine if spiral arms could be clearly discerned. It was
found that the highest density of points in a position-position plot
approximately follows the spiral arms proposed by TC93; however, 
the scatter about the spiral arm centroids is about as
large as the separation between the arms of TC93 making the distinction
between spiral arms indistinct. Positions of the H {\small II} regions in an l$-$v plot rely on observed
quantities alone and are independent of Galactic rotation models. Comparing the locus of points of the TC93 spiral arms with the distribution of H {\small II} regions whose distance ambiguities have been resolved shows an increase in the density of sources at the points where the loci of spiral arms 
are approaching the locus of tangent point velocities and a lower density between the arm loci. 
However, even in an l$-$v plot, the scatter is too large to clearly trace spiral arms over significant
segments in Galactic longitude. Four reasons were discussed why H {\small II}
regions do not seem to clearly delineate spiral arms in our Galaxy.
They are: 1) spiral arms have intrinsic widths both in space and velocity
which will smear the distribution of sources about arm centroids; 2)
peculiar motions (i.e., departures from the motions included in the
rotation model from which distances were determined) will smear out the
distribution around a spiral arm centroid even further; 3) a limited
number of discrete H {\small II} regions do not provide the velocity continuity
necessary to recognize large scale kinematic features; and, 4) errors in
the near/far distance ambiguity assignment can cause large distance
errors which will further confuse a position-position plot. Thus, we
conclude that without a very large number of H {\small II} regions whose distance
ambiguities have been resolved in combination with more sophisticated
Galactic rotation models that take into account non-circular motions both
in the plane and perpendicular to it, it is unlikely that a more continuous spiral pattern will emerge from kinematic studies of H {\small II}
regions. Resolution of distance ambiguities of
H {\small II} regions is not a fruitless exercise, however. Accurate distances, which was the
main motivation for undertaking this study, are crucial for the
determination of physical properties (e.g., absolute sizes, luminosities,
and masses) of massive star formation regions.

\acknowledgments
We thank F.D. Ghigo and R.C. Bignell for help during the observations and data reduction. We express our gratitude to Butler Burton for his thought-provoking comments on the manuscript that significantly improved this paper. We also thank him for providing the HI data that we used in our analysis. MS gratefully acknowledges support of a GBT Student Support Fellowship. EBC acknowledges partial
support from NSF grant AST-0303689. PH acknowledges partial support from Research Corporation award CC4996 and NSF grant AST-0098524. SK acknowledges partial support from CONACyT grant E-36568 and DGAPA-UNAM grant IN118401.

\begin{figure}
\figurenum{1}
\plotone{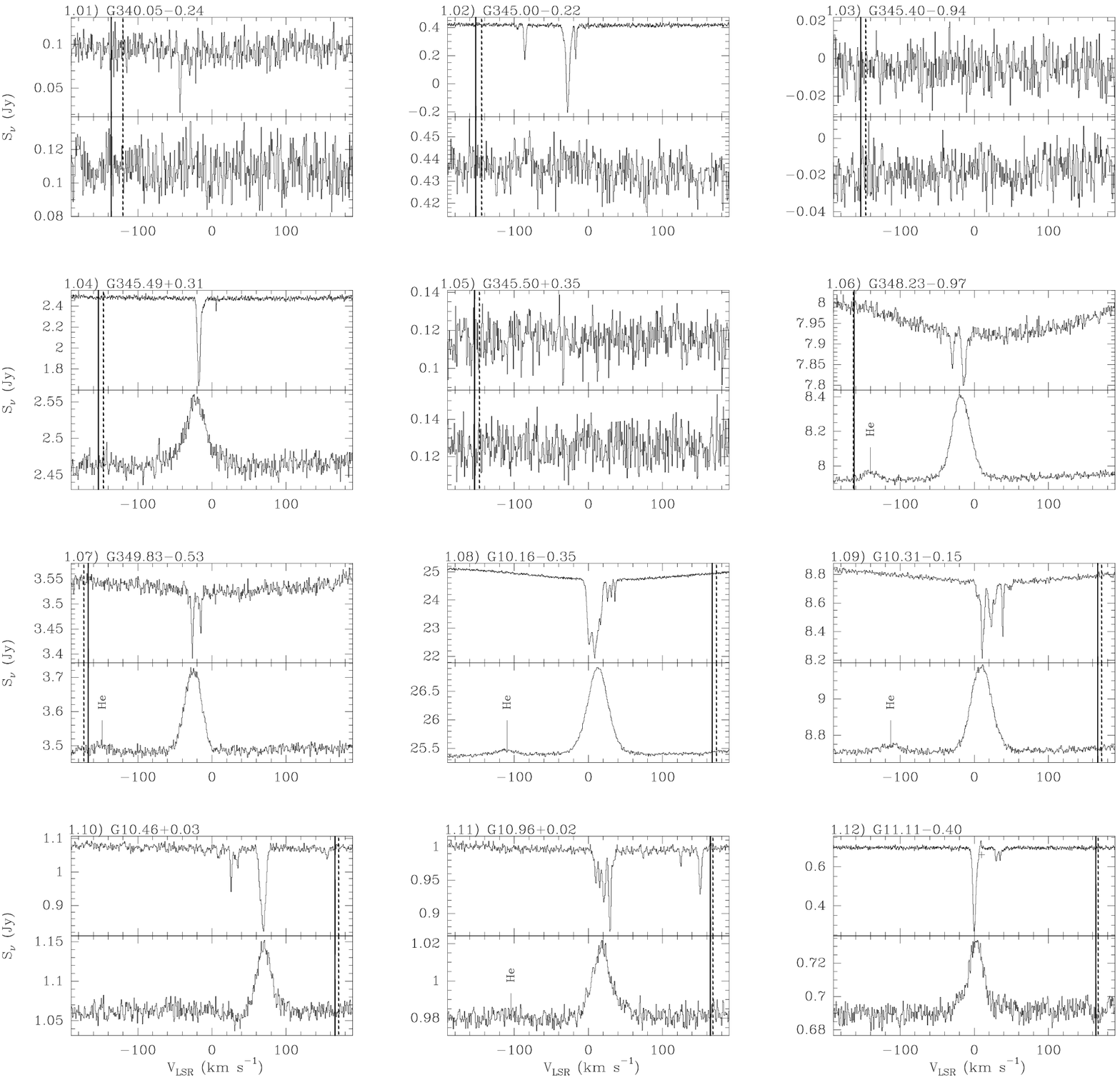}
\vspace*{-3.5cm}
\caption{Spectra of H$_2$CO ({\it upper panel}) and H110$\alpha$ ({\it lower
panel}) observed toward UC H {\small II} regions. The solid and dashed vertical lines
indicate the tangential velocity according to the Brand \& Blitz (1993)
and Clemens (1985) models of Galactic rotation, respectively. Absorption in the 
OFF-position is indicated by a plus sign. Tentative detections of He and C recombination 
lines are indicated explicitly. \label{spectra}}
\end{figure}

\begin{figure}
\figurenum{1}
\plotone{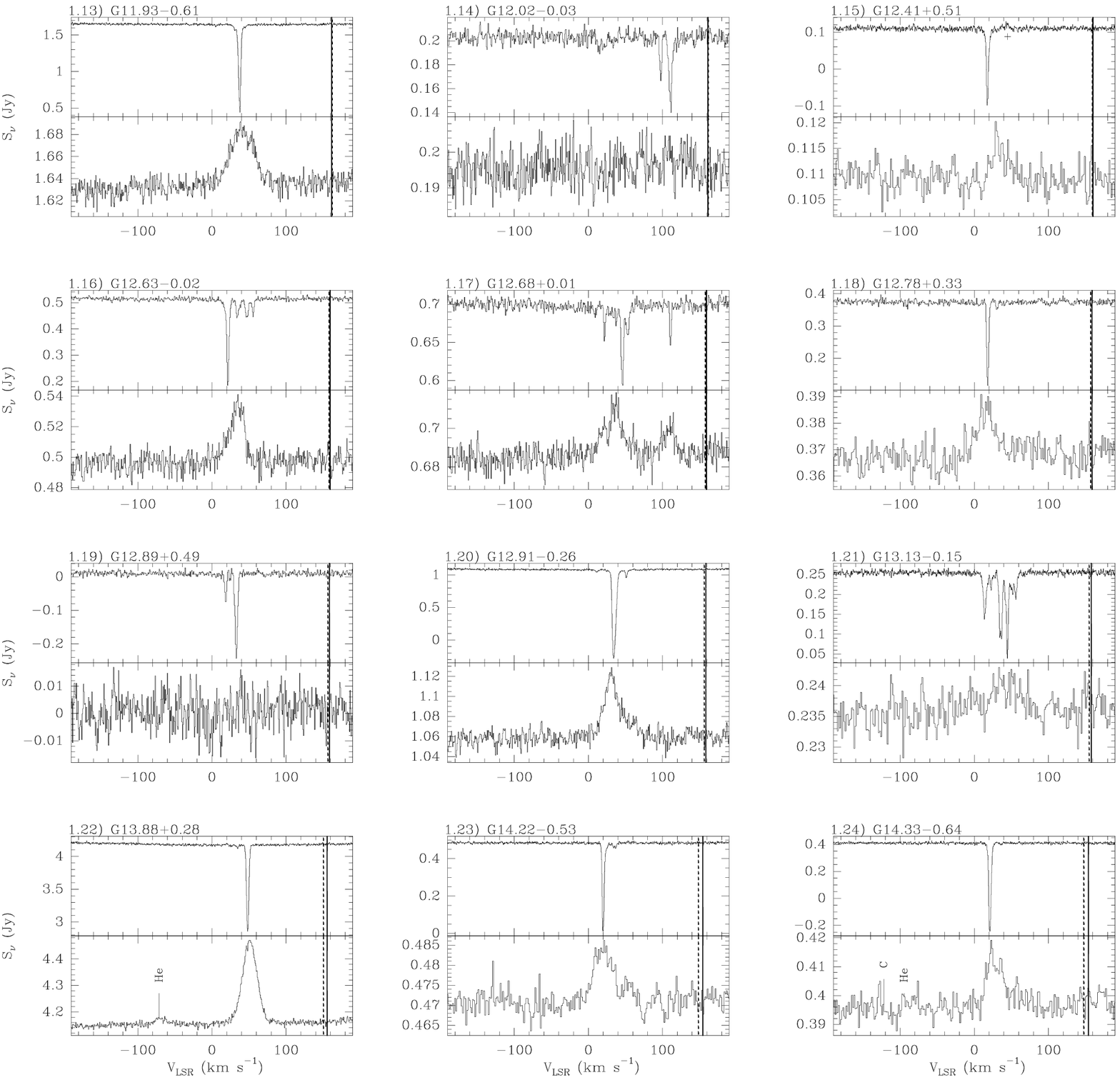}

\end{figure}

\begin{figure}
\figurenum{1}
\epsscale{0.9}
\plotone{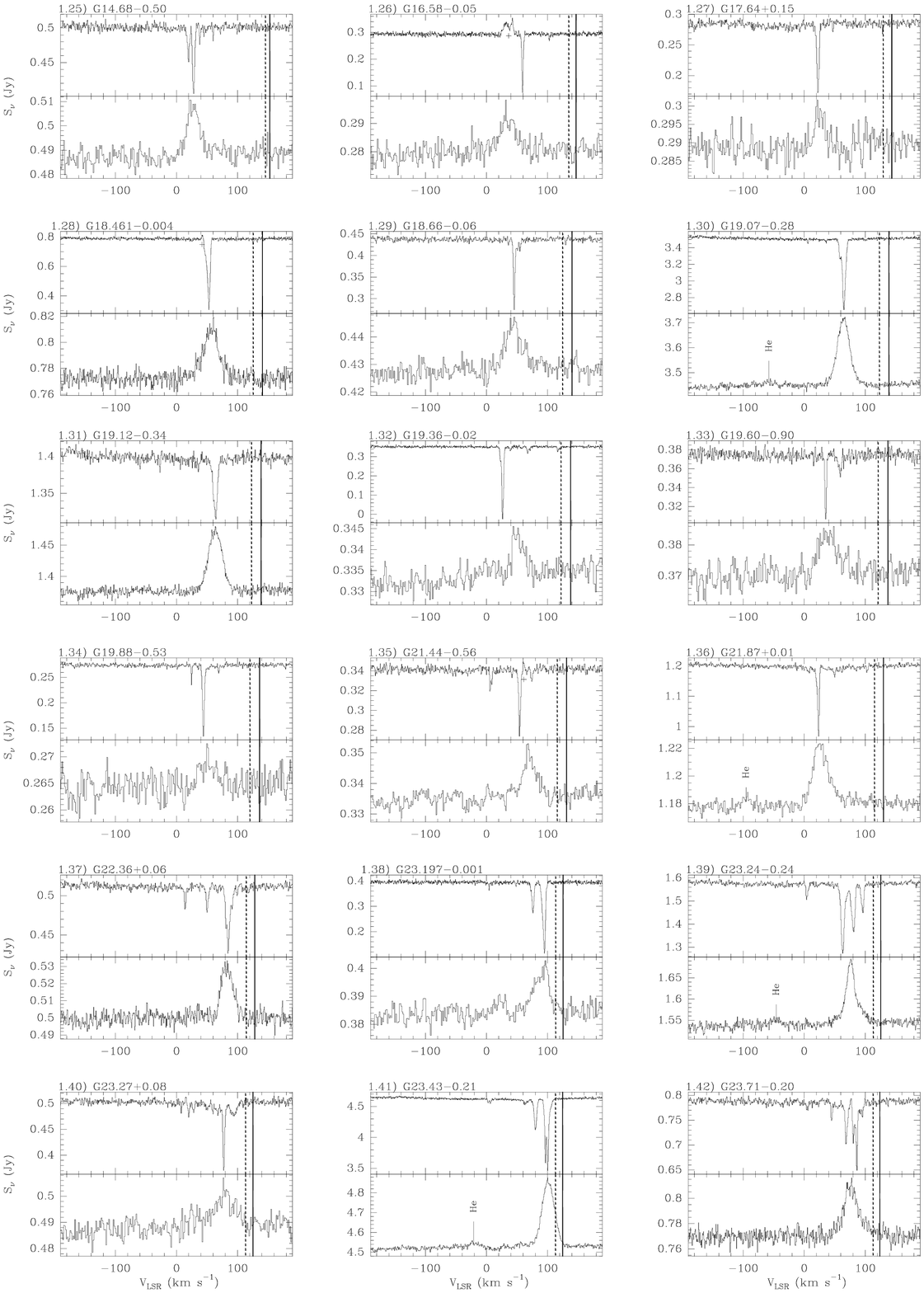}

\end{figure}

\begin{figure}
\figurenum{1}
\epsscale{0.9}
\plotone{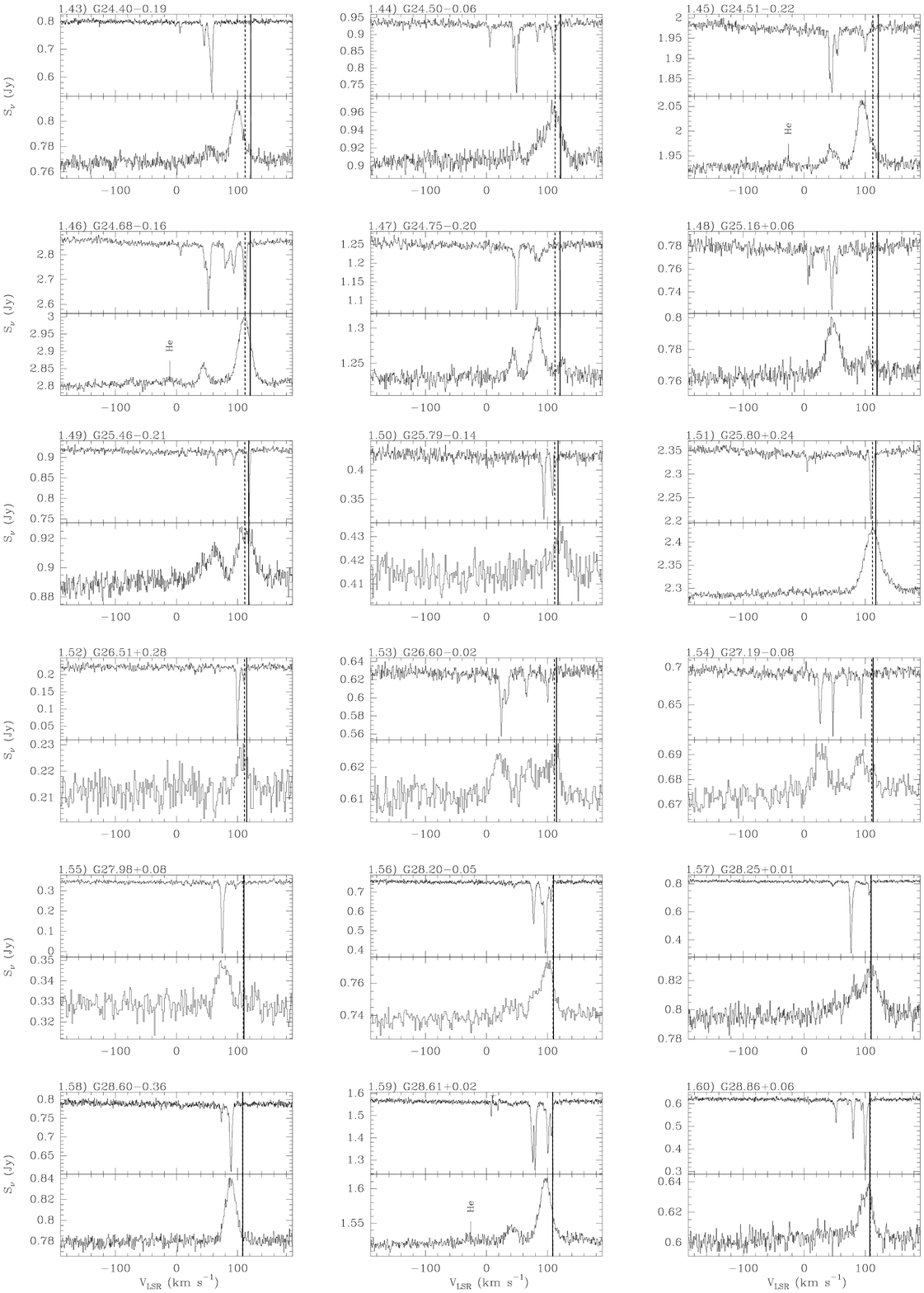}

\end{figure}

\begin{figure}
\figurenum{1}
\plotone{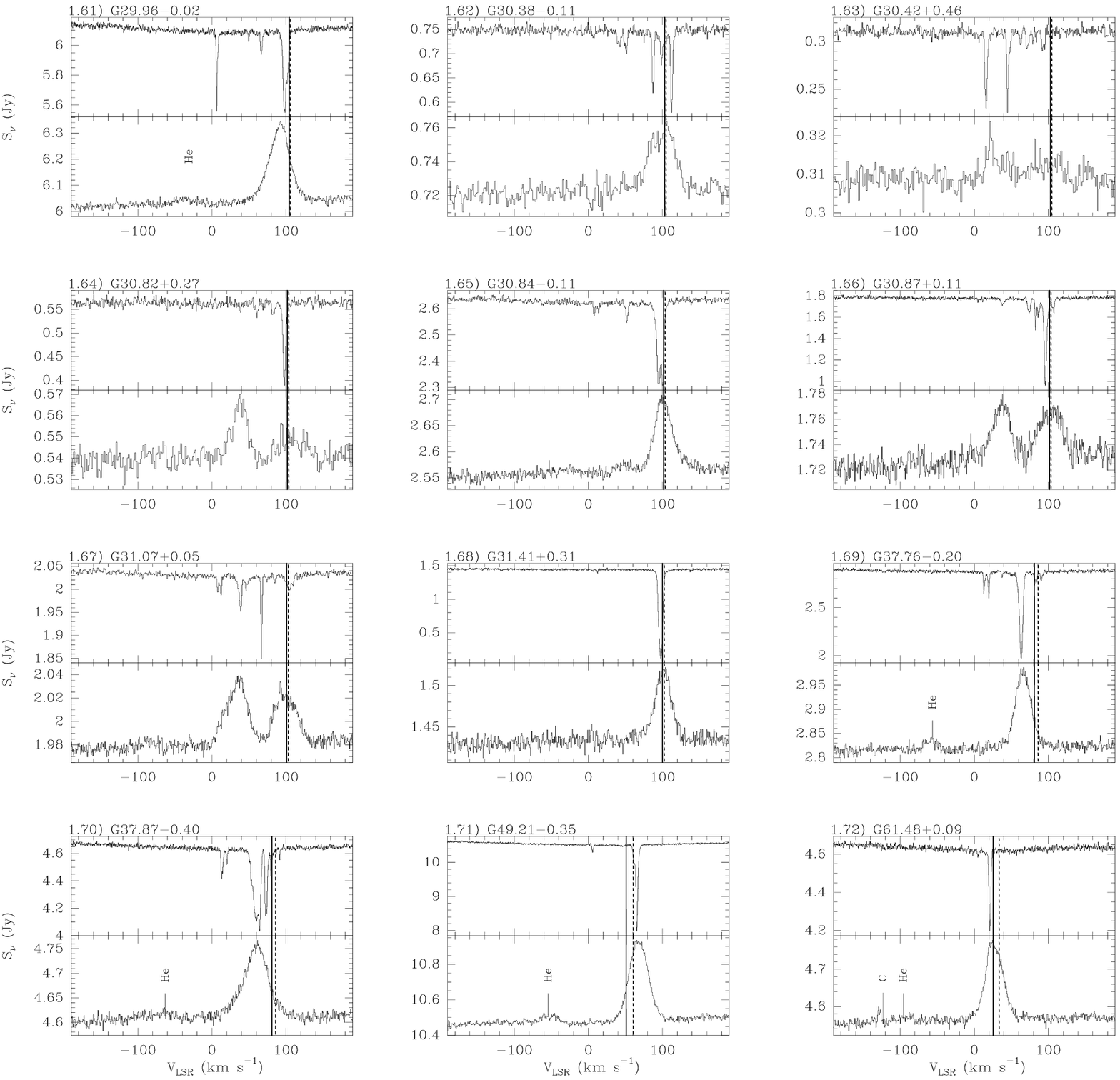}
\vspace*{-3.5cm}

\end{figure}

\begin{figure}
\figurenum{2}
\plotone{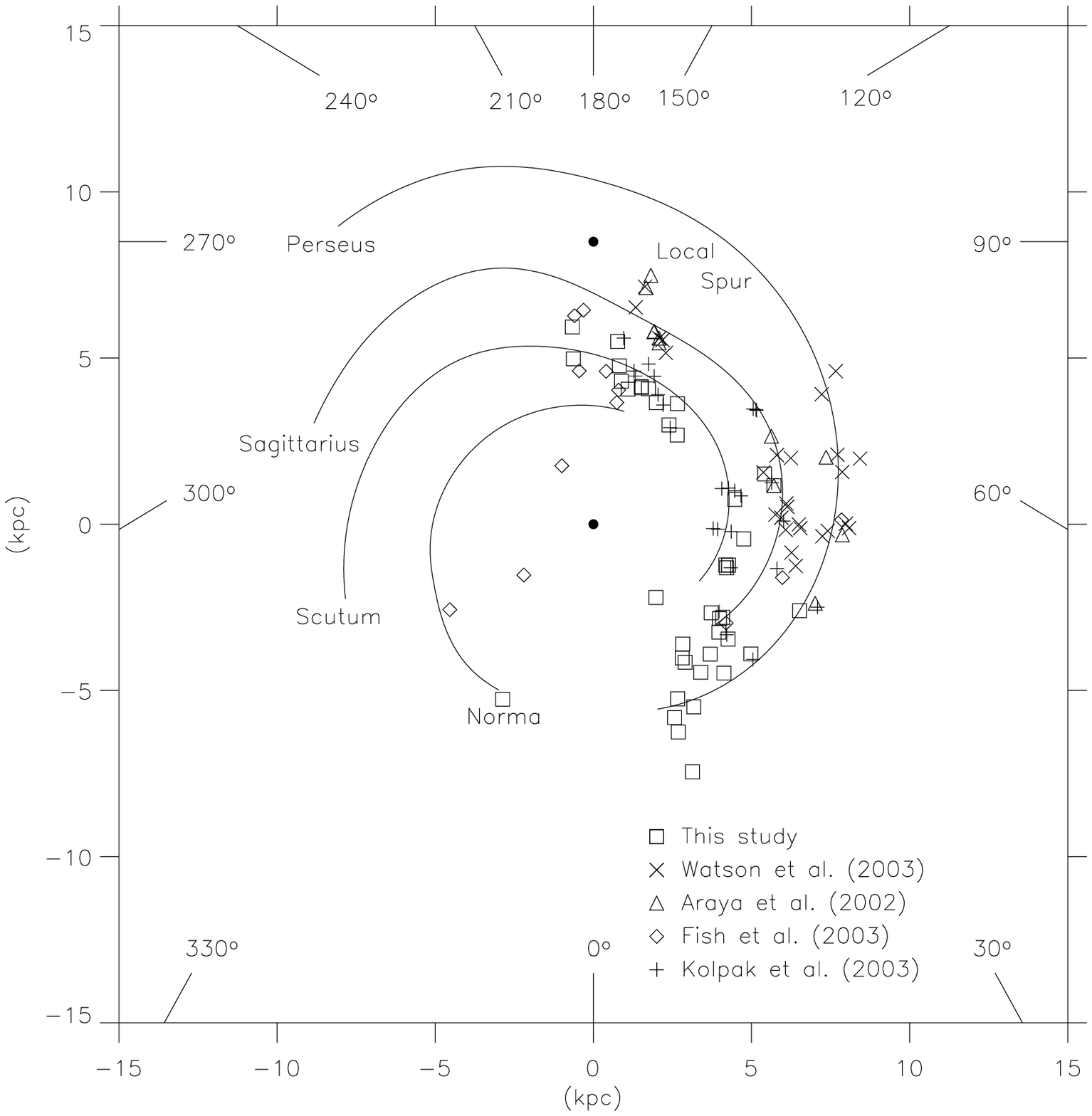}
\caption{The distribution of H {\small II} regions projected onto the Galactic plane from several H$_{2}$CO and H {\small I} absorption line surveys which resolve the distance ambiguity (sources located at the tangent point are not included). The spiral arm model by Georgelin and Georgelin (1976) modified by Taylor and Cordes (1993) is indicated by solid lines. The Galactic center and position of the Sun are indicated by filled circles.}
\end{figure}

\begin{figure}
\figurenum{3}
\plotone{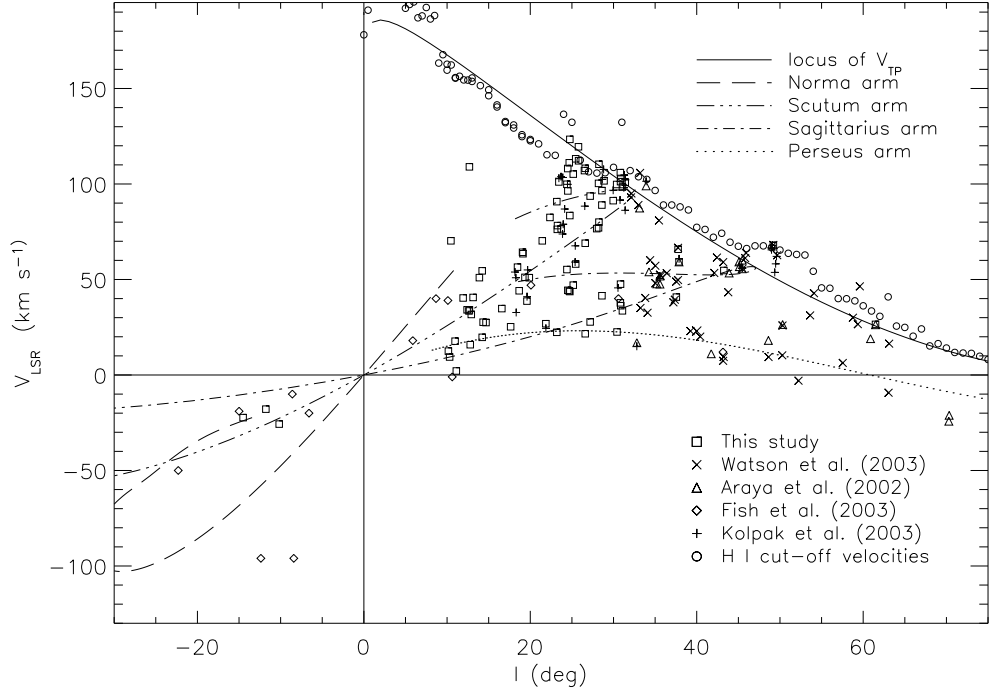}
\caption{The distribution of all the observed H {\small II} regions in the same surveys as in Figure 2 in the longitude-velocity domain. The locus of tangential velocity (V$_{TP}$) for the rotation curve of Brand and Blitz (1993) is the outer envelope to the distribution that would be expected in the absence of any random motions. The Taylor and Cordes (1993) model of four spiral arms is also shown. 
The H {\small I} cut-off velocities are taken from Westerhout and Wendlandt (1982; l = 10$\arcdeg$ - 20$\arcdeg$), Burton and Liszt (1983; l = 0$\arcdeg$ - 13$\arcdeg$), and Fich et al. (1989; l = 15$\arcdeg$ - 90$\arcdeg$). The Westerhout and Wendlandt (1982) and Burton and Liszt (1983) data were kindly provided by B. Burton.}
\end{figure}

\clearpage

\begin{deluxetable}{p{3.5cm}lllrc}
\tablecaption{Observed Sources}
\tabletypesize{\scriptsize}
\tablewidth{0pt}

\tablehead{

\colhead{Source}&
\colhead{$\alpha$(B1950)}&
\colhead{$\delta$(B1950)}&
\colhead{IRAS name}&
\colhead{S$_{100\mu m}$} &
\colhead{t$_{int}^{\tablenotemark{a}}$ } \\

\colhead{} &
\colhead{($^{h\phm{66}m\phm{66}s}$)}&
\colhead{($^{\circ}\phm{66}'\phm{66}''$)} &
\colhead{} &
\colhead{(10$^{3}$ Jy)} &
\colhead{ (min.)}
}

\startdata

G340.05$-$0.24 \dotfill & 16 44 34.0 & $-$45 16 18 & 16445$-$4516 & 8.54 & 10\\
G345.00$-$0.22 \dotfill & 17 01 38.8 & $-$41 24 59 & 17016$-$4124 & 7.03 & 22\\
G345.40$-$0.94 \dotfill & 17 05 59.8 & $-$41 32 09 & 17059$-$4132 & 25.8 & 10\\
G345.49$+$0.31 \dotfill & 17 00 59.9 &  $-$40 42 19 & 17009$-$4042 &  18.20 & 10\\
G345.50$+$0.35 \dotfill & 17 00 51.0 & $-$40 40 18 & 17008$-$4040 & 18.20 & 10\\
G348.23$-$0.97 \dotfill & 17 14 56.4 &  $-$39 16 02 & 17149$-$3916 &  9.16 & 10\\
G349.83$-$0.53 \dotfill & 17 17 48.6 &  $-$37 42 22 & 17178$-$3742 &  6.88 & 10\\
G10.16$-$0.35  \dotfill & 18 06 25.9 &  $-$20 20 04 & 18064$-$2020 & 27.30 & 12\\
\enddata 
\tablenotetext{a}{The sum of the ON-Source and OFF-Source integration times.}

\end{deluxetable}

\begin{deluxetable}{p{3.5cm}ccc}
\tablecaption{Velocity Resolutions of the Spectra\tablenotemark{a}}
\tabletypesize{\scriptsize}
\tablewidth{0pt}

\tablehead{

\colhead{Source} &
\colhead{$\Delta$V$_{H_{2} CO}$} &
\colhead{$\Delta$V$_{H110 \alpha}$} &
\colhead{Fig.} \\

\colhead{} &
\colhead{(km s$^{-1}$)} & 
\colhead{(km s$^{-1}$)} &
\colhead{}

}

\startdata

G340.05$-$0.24  \dotfill     & 0.76 & 0.76 & 1.01 \\
G345.00$-$0.22  \dotfill     & 0.38 & 0.76 & 1.02 \\
G345.40$-$0.94  \dotfill     & 0.76 & 0.76 & 1.03 \\
G345.49$+$0.31  \dotfill     & 0.38 & 0.76 & 1.04 \\
G345.50$+$0.35  \dotfill     & 0.76 & 0.76 & 1.05 \\
G348.23$-$0.97  \dotfill     & 0.76 & 0.76 & 1.06 \\
G349.83$-$0.53  \dotfill     & 0.76 & 0.76 & 1.07 \\
G10.16$-$0.35   \dotfill     & 0.38 & 0.76 & 1.08 \\
\enddata

\tablenotetext{a}{The intrinsic velocity resolution of the spectra is 0.188~km~s$^{-1}$. One, two, and three Hanning smoothings reduce it to 0.38~km~s$^{-1}$, 0.76~km~s$^{-1}$, and 1.50~km~s$^{-1}$, respectively.}

\end{deluxetable}

\begin{deluxetable}{p{3.5cm}rrrr}
\tablecaption{H110$\alpha$ Line Parameters of Sources with a Single Velocity Component}
\tabletypesize{\footnotesize}
\tablewidth{0pt}

\tablehead{

\colhead{Source}&
\colhead{S$_{H110\alpha}$}&
\colhead{V$_{LSR}$} &
\colhead{FWHM}&
\colhead{$\int$S$_v$ dv}\\

\colhead{} &
\colhead{(mJy)} &
\colhead{(km s$^{-1}$)} &
\colhead{(km s$^{-1}$)} &
\colhead{(Jy km s$^{-1}$)}

}

\startdata

G340.05$-$0.24  \dotfill&   \nodata      &  \nodata      &  \nodata      & \nodata    \\
G345.00$-$0.22  \dotfill&   \nodata      &  \nodata      &  \nodata      & \nodata    \\		   
G345.40$-$0.94  \dotfill&   \nodata      &  \nodata      &  \nodata      & \nodata    \\		   
G345.49$+$0.31  \dotfill&   84.8 (6.0)   &  -22.3 (0.5)  &   33.8 ( 1.3) & 3.05 (0.09) \\
G345.50$+$0.35  \dotfill&   \nodata      &  \nodata      &  \nodata      & \nodata    \\		   
G348.23$-$0.97  \dotfill&   471.7 (10.4) &  -17.9 (0.1)  &  29.0 (0.2)   & 14.54 (0.10) \\	   
G349.83$-$0.53  \dotfill&   238.3 (7.7)  &  -25.7 (0.1)  &  25.9 (0.3)   & 6.57 (0.06)  \\		   
G10.16$-$0.35   \dotfill&   1493.7 (26.3)&  12.5 (0.1)   &  32.9 (0.1)   &  52.35 (0.19) \\
\enddata

\tablecomments{The errors quoted in paranthesis are the formal 1$\sigma$ errors of the fits.}

\end{deluxetable}

\begin{deluxetable}{p{3.5cm}rrrr}
\tablecaption{H110$\alpha$ Line Parameters of Sources with Multiple H110$\alpha$ Velocity Components }
\tabletypesize{\footnotesize}
\tablewidth{0pt}

\tablehead{

\colhead{Source}&
\colhead{S$_{H110\alpha}$}&
\colhead{V$_{LSR}$} &
\colhead{FWHM}&
\colhead{$\int$S$_v$ dv}\\

\colhead{} &
\colhead{(mJy)} &
\colhead{(km s$^{-1}$)} &
\colhead{(km s$^{-1}$)} &
\colhead{(Jy km s$^{-1}$)}

}

\startdata

G12.68$+$0.01 H1                   \dotfill&  21.3 (4.7)   & 34.1 (0.7) &  30.4 (1.7) & 0.69 (0.03) \\ 
G12.68$+$0.01 H2\tablenotemark{a}  \dotfill&  \nodata      &109.0 (0.9) &  \nodata    & \nodata    \\
                                           &               &            &             &             \\
G14.22$-$0.53 H1                   \dotfill&  13.5 (3.4)   & 19.7 (0.8) &  28.4 (2.0) & 0.41 (0.02) \\
G14.22$-$0.53 H2\tablenotemark{a}  \dotfill&  \nodata      & 54.5 (1.7) &  \nodata    & \nodata     \\
                                           &               &            &             &             \\
G23.20$-$0.00   H1                 \dotfill&  14.3 (3.1)   & 90.7 (0.7) &  28.3 (1.6) & 0.43 (0.02) \\ 
G23.20$-$0.00  H2\tablenotemark{a}\dotfill&  \nodata      & 22.4 (1.7) &  \nodata    & \nodata     \\
\enddata

\tablenotetext{a}{The intensity of the second line is less than 3$\sigma$; only its position is given.}

\end{deluxetable}

\begin{deluxetable}{p{4.5cm}rrrrr}
\tablecaption{H$_2$CO Line Parameters}
\tabletypesize{\footnotesize} 
\tablewidth{0pt}

\tablehead{ 

\colhead{Source} &
\colhead{S$_{H_2CO}$}&
\colhead{V$_{LSR}$} &
\colhead{FWHM} &
\colhead{$\int$S$_v$ dv}\\

\colhead{Direction\tablenotemark{a}} &
\colhead{(mJy)} &
\colhead{(km s$^{-1}$)} &
\colhead{(km s$^{-1}$)}&
\colhead{(Jy km s$^{-1}$)}

}

\startdata 

G340.05$-$0.24 MC 1 \dotfill&  $-$67.6 (11.2)  & $-$43.0 (0.2)    & 2.5 (0.4)  & $-$0.18 (0.02) \\
                            &                  &                  &            &                \\       
G345.00$-$0.22 MC 1 \dotfill&  $-$571.0 (35.4) & $-$27.7 ($<$0.1) & 5.9 (0.1)  & $-$3.56 (0.04) \\
G345.00$-$0.22 MC 2 \dotfill&  $-$238.4 (15.1) & $-$85.5 ($<$0.1) & 3.0 (0.1)  & $-$0.75 (0.01) \\
G345.00$-$0.22 MC 3 \dotfill&  $-$234.4 (16.2) & $-$17.0 ($<$0.1) & 2.2 (0.1)  & $-$0.56 (0.02) \\
G345.00$-$0.22 MC 4 \dotfill&  $-$28.8 (9.1 )  & $-$95.5 (0.2)    & 2.6 (0.3)  & $-$0.08 (0.01) \\
                            &                  &                  &            &                \\    
G345.40$-$0.94 MC 1 \dotfill&  \nodata         & \nodata          & \nodata    & \nodata        \\
\enddata

\tablenotetext{a}{Indicates the direction of the H$_2$CO cloud, but
we do not imply an association between the H$_2$CO cloud and the background H {\scriptsize II}
region.}

\end{deluxetable}

\begin{deluxetable}{ll}
\tablecaption{Criteria for Resolution of Distance Ambiguity}
\tabletypesize{\small}
\tablewidth{0pt}

\tablehead{

\colhead{Distance Classification} &
\colhead{Classification Criteria}

}

\startdata
\multicolumn{2}{l}{\bf H {\small II} Regions} \\

Far &  $\rm \exists~~ |V_{H_2CO}|:~ (|V_{HII}|+10~km~s^{-1}) < |V_{H_2CO}| <
(|V_{TP}|+10~km~s^{-1}) $ \\

& and~~ ($|$V$_{HII}|$ $+$ 10~km~s$^{-1}$) $<$ $|$V$_{TP}|$ \\

Near & $\rm \nexists~~ |V_{H_2CO}|:~ (|V_{HII}|+10~km~s^{-1}) < |V_{H_2CO}| <
(|V_{TP}|+10~km~s^{-1}) $\\

& and~~ ($|$V$_{HII}|$ $+$ 10~km~s$^{-1}$) $<$ $|$V$_{TP}|$ \\

No classification \tablenotemark{a} & ($|$V$_{TP}$$|$ $-$ 10 km~s$^{-1}$) $<$ $|$V$_{HII}$$|$ $<$ $|$V$_{TP}$$|$ \\

Tangent Point \tablenotemark{b}  & $|$V$_{HII}$$|$ $\geq$ $|$V$_{TP}|$ \\

 & \\
& \\

\multicolumn{2}{l}{\bf H$_{2}$CO Clouds}  \\

Near & $|$V$_{H_{2}CO}|$ $<$ ($|$V$_{HII}|$ $-$ 10 km~s$^{-1}$) \\

Associated with an H{\small II} region & ($|$V$_{HII}|$ $-$ 10 km~s$^{-1}$) $<$ $|$V$_{H_{2}CO}|$ $<$ ($|$V$_{HII}|$ $+$ 10 km~s$^{-1}$)  \\

Ambiguous  & ($|$V$_{HII}|$ $+$ 10 km~s$^{-1}$) $<$ $|$V$_{H_{2}CO}|$ $<$ ($|$V$_{TP}|$ $-$ 10 km~s$^{-1}$) \\

No classification \tablenotemark{a} &($|$V$_{TP}|$ $-$ 10 km~s$^{-1}$) $<$ $|$V$_{H_{2}CO}|$ $<$ $|$V$_{TP}|$ \\

Tangent Point \tablenotemark{b} & $|$V$_{H_{2}CO}|$ $\geq$ $|$V$_{TP}|$ \\

\enddata

\tablenotetext{a}{We do not give a classification or distance, because we cannot unambiguously distinguish
between the near and far position of a source in this velocity range.}

\tablenotetext{b}{We locate a source at the tangent point, but
this position should be taken with caution since the rotation model does not predict a position for these velocities.}

\end{deluxetable}

\begin{deluxetable}{p{3cm}rccccrrcc}
\tablecaption{Kinematic Parameters}
\tabletypesize{\footnotesize}
\rotate
\tablewidth{0pt}

\tablehead{

\colhead{Source}	&
\colhead{V$_{LSR}$}	&
\colhead{D$_{near}$}	&
\colhead{D$_{far}$}&
\colhead{V$_{TP}$}	&
\colhead{D$_{TP}$}	&
\colhead{D$_{LSR}$} &
\colhead{z}&
\colhead{R$_{GC}$}	&
\colhead{Near/Far \tablenotemark{1}} 
\\

\colhead{} &
\colhead{(km s$^{-1}$)}	&
\colhead{(kpc)}	&
\colhead{(kpc)}	&
\colhead{(km s$^{-1}$)}&
\colhead{(kpc)}	&
\colhead{(kpc)}	   &
\colhead{(pc)} 	&
\colhead{(kpc)} &
\colhead{} 

}

\startdata

G345.49$+$0.31      \dotfill  &  $-$22.3 (0.5)  &  2.7   &     13.8   & $-$153.5 & 8.2  & 2.7$^{+1.0}_{-0.8}$  &   14.4  & 6.0  &  N   \\ 			    
MC 1                \dotfill  &  $-$11.7 (0.2)  &  1.6   &     14.9   & $-$153.5 & 8.2  & 1.6$^{+1.3}_{-1.0}$  &   8.7  & 7.0  &  N   \\ 			   
MC 2                \dotfill  &     6.10 (0.03) & \nodata&   \nodata  & $-$153.5 & 8.2  & 17.6$^{+1.7}_{-2.6}$ &   95.2  & 9.6  &  F   \\   
&&&&&&&&&\\
G348.23$-$0.97      \dotfill  &  $-$17.9 (0.1)  &  2.6   &     14.1   & $-$162.1 & 8.3  & 14.1$^{+0.9}_{-1.2}$ &  $-$238.0  & 6.0  &  F  \\ 			  
MC 1                \dotfill  &  $-$14.2 (0.1)  &  2.2   &     14.5   & $-$162.1 & 8.3  & 14.5$^{+1.0}_{-1.4}$ &  $-$244.9  & 6.4  &  FS  \\ 			  
&&&&&&&&&\\
G349.83$-$0.53      \dotfill  &  $-$25.7 (0.1)  &  3.6   &     13.1   & $-$167.0 & 8.4  & 3.6$^{+1.0}_{-0.7}$  &  $-$33.1  & 5.0  &  N \\
\enddata

\tablenotetext{1}{N - a source at the near kinematic distance; F - a source at the far distance; NS and FS - a molecular cloud associated with an UC H {\small II} region 
(i.e., V$_{H_2 CO}$ within 10 km~s$^{-1}$ of V$_{H II}$) at the near and far distance, respectively.}

\tablenotetext{a}{No classification or distance is given for this source; see footnotes to Table 6.}

\tablenotetext{b}{This source is located at the tangent point; see footnotes to Table 6.}

\end{deluxetable}

\end{document}